\documentclass[aps,pre,twocolumn,groupedaddress,superscriptaddress,showpacs]{revtex4-1}

\usepackage{graphicx}
\usepackage{color}
\usepackage{amsmath}
\usepackage{amsfonts}
\usepackage{amssymb}
\usepackage{amsmath}
\usepackage{dcolumn}

\begin{document}
\title{Fracturing ranked surfaces}
  \author{K. J. Schrenk}
    \affiliation{Computational Physics for Engineering Materials, IfB, ETH Zurich, Schafmattstrasse 6, CH-8093 Zurich, Switzerland}

  \author{N. A. M. Ara\'ujo}
    \email{nuno@ethz.ch}
    \affiliation{Computational Physics for Engineering Materials, IfB, ETH Zurich, Schafmattstrasse 6, CH-8093 Zurich, Switzerland}

  \author{J. S. Andrade, Jr.}
    \affiliation{Computational Physics for Engineering Materials, IfB, ETH Zurich, Schafmattstrasse 6, CH-8093 Zurich, Switzerland}
    \affiliation{Departamento de F\'isica, Universidade Federal do Cear\'a, 60451-970 Fortaleza, Cear\'a, Brazil}

  \author{H. J. Herrmann}
    \affiliation{Computational Physics for Engineering Materials, IfB, ETH Zurich, Schafmattstrasse 6, CH-8093 Zurich, Switzerland}
    \affiliation{Departamento de F\'isica, Universidade Federal do Cear\'a, 60451-970 Fortaleza, Cear\'a, Brazil}
\begin{abstract}
Discretized landscapes can be mapped onto ranked surfaces, where every
element (site or bond) has a unique rank associated with its
corresponding relative height. By sequentially allocating these elements
according to their ranks and systematically preventing the occupation of
bridges, namely elements that, if occupied, would provide global
connectivity, we disclose that bridges hide a new tricritical point at
an occupation fraction $p=p_{c}$, where $p_{c}$ is the percolation
threshold of random percolation. For any value of $p$ in the interval
$p_{c}< p \leq 1$, our results show that the set of bridges has a
fractal dimension $d_{BB} \approx 1.22$ in two dimensions. In the limit
$p \rightarrow 1$, a self-similar fracture is revealed as a singly
connected line that divides the system in two domains. We then unveil
how several seemingly unrelated physical models tumble into the same
universality class and also present results for higher dimensions.
\end{abstract}%
\maketitle
\begin{figure*}
      \begin{center}
        \includegraphics[width=0.8\textwidth]{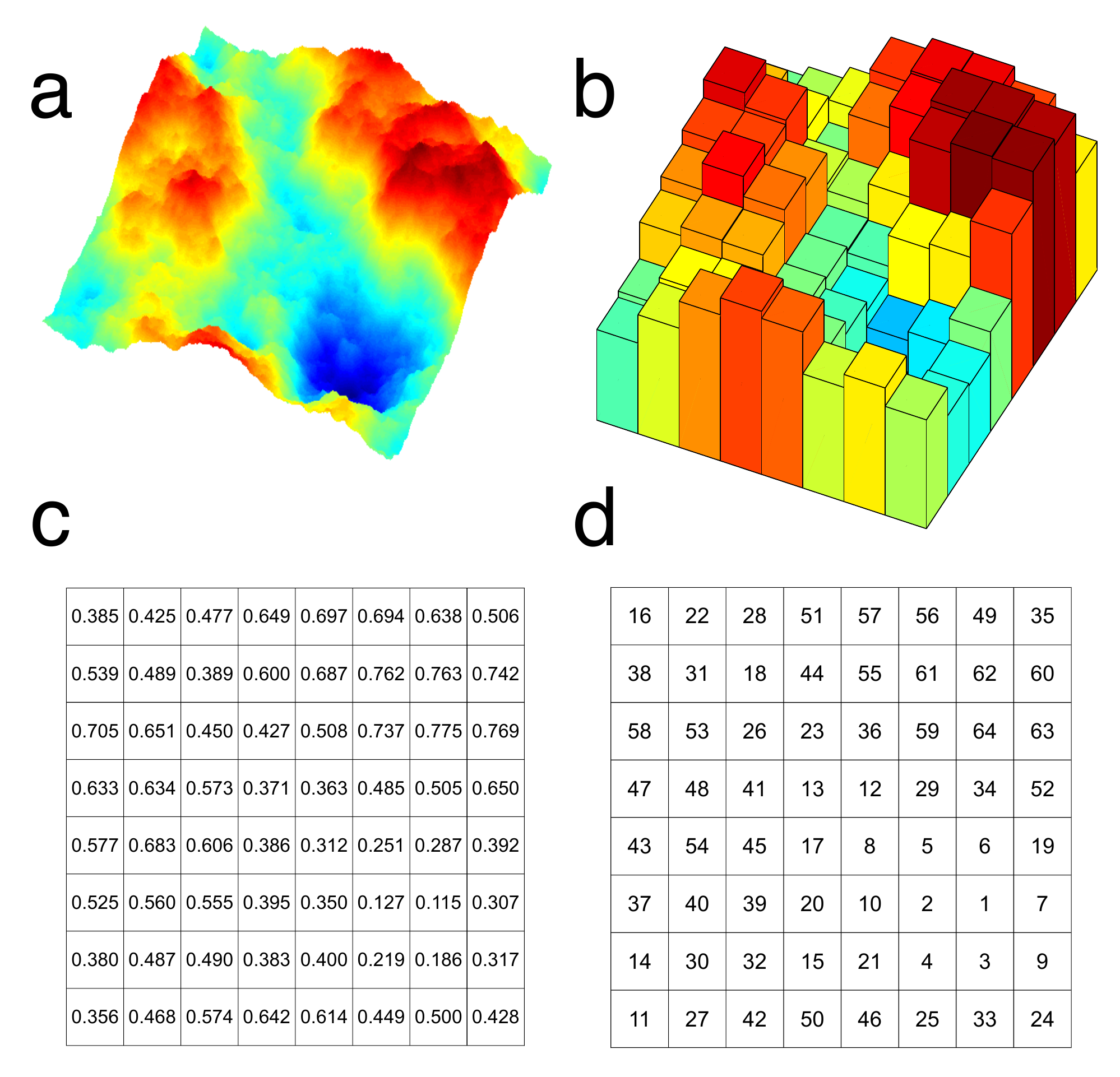}
      \end{center}
      \caption{
        The generation process of a ranked surface.
        The landscape in (a) is coarse-grained to the low-resolution 
        system of $8 \times 8$ shown in (b), and then represented 
        as a discretized map of local heights, as depicted in (c). 
        By ranking these heights in crescent order, one obtains the 
        ranked surface in (d). In fact, the landscape shown in (a)
        is a high resolution synthetic map obtained from a fractional 
        Brownian motion simulation based on the Fourier filtering method
        \cite{Prakash92,Sahimi94,Sahimi96,Makse96,Peitgen88,Mandelbrot68,Oliveira11}.
        \label{fig::rank}}
\end{figure*}

Any real landscape can be duly coarse-grained and represented as a
two-dimensional {\it discretized map} of regular cells (e.g., a square
lattice of sites or bonds) to which average heights can be associated.
This process is exemplarily shown in Figs.~\ref{fig::rank}(a)-(c). As such, the concept
of discretized maps has been considered as a way to delimit spatial
boundaries in a wide range of seemingly unrelated problems, ranging from
tracing water basins and river networks in landscapes
\cite{Stark91,Maritan96,Manna96,Knecht12,Baek11} to the identification
of cancerous cells in human tissues \cite{Yan06,Ikedo07}, and the study
of spatial competition in multispecies ecosystems
\cite{Kerr06,Mathiesen11}. Moreover, previous studies have shown that
cracks or surviving paths through discretized maps possess a universal
fractal dimension which can be physically realized in terms of optimal
paths under strong disorder \cite{Cieplak94,Cieplak96,Fehr09,Fehr11},
optimal path cracks \cite{Andrade09,Oliveira11}, loopless percolation
\cite{Porto97,Porto99}, or minimum spanning trees
\cite{Barabasi96,Dobrin01,Jackson10,Ciftci11,Goyal11,Hubbe11}.  Here we
show that all these problems can be understood in terms of the same
universal concept of {\it fracturing a ranked surface}.

We start by defining a ranked surface. Given a two-dimensional
discretized map of size $L \times L$, we generate a list containing the
heights of its elements (sites or bonds) in crescent order, and then
replace the numerical values in the original map by their corresponding
ranks. As depicted in Fig.~\ref{fig::rank}(d), the result is a ranked
surface. The process of fracture generation is rather simple. Once the
ranked surface is obtained, we sequentially occupy the elements of an
empty lattice with the same size following the crescent rank order of
the corresponding elements (i.e., in the same position) on the ranked
surface. During each step of the allocation process, only bridges,
identified as those lattice elements which, once occupied, would create
a spanning cluster (i.e., a globally connecting cluster)
\cite{Stauffer94}, are never occupied. These elements will eventually form
a macroscopic fracture.

\begin{figure*}
      \begin{center}
        \includegraphics[width=\textwidth]{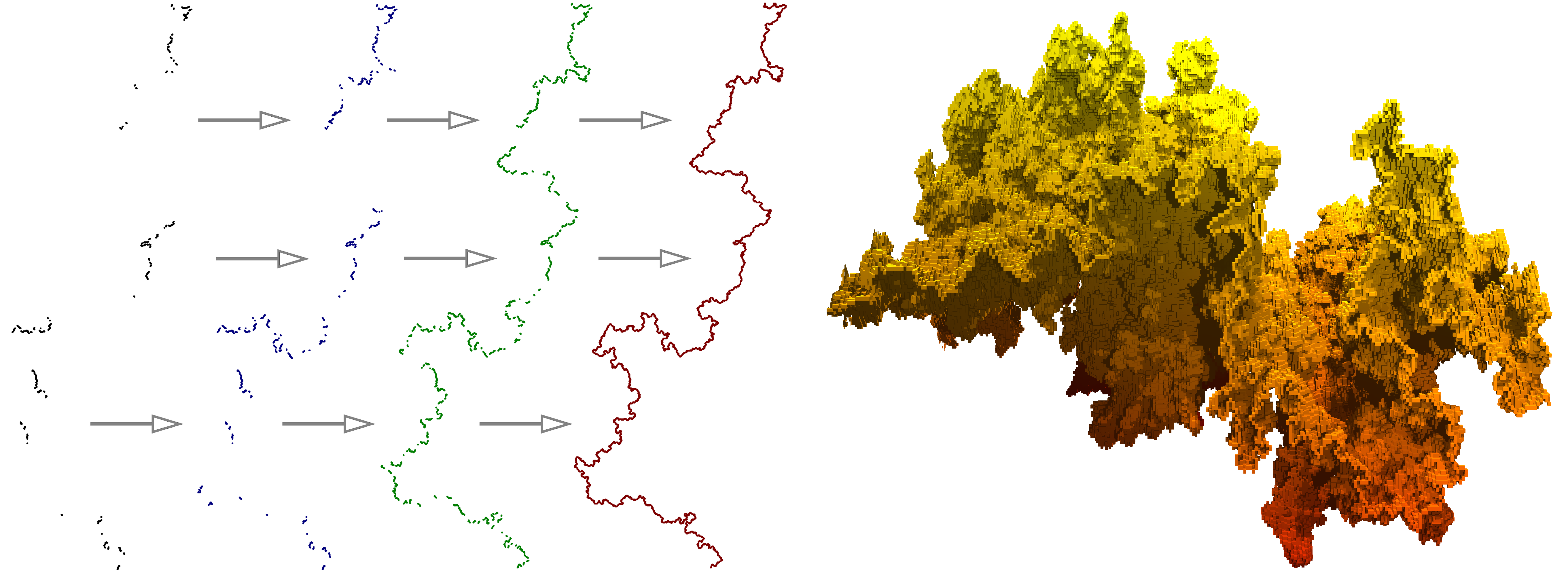}
      \end{center}
      \caption{
        Snapshots of the fractal set of bridges in two (line)
and three (surface) dimensions.
        For $2D$ four stages are seen (from left to right): $p=p_c$ (black), $p=1.01p_c$ (blue), $p=1.05p_c$ (green), and $p=1$ (red), while for $3D$ only the final set is shown.
        We considered in $2D$ a square lattice with $1024^2$ sites and, in $3D$, a simple-cubic lattice with $512^3$ sites.
        The fractal dimension is $d_{BB}=1.215\pm0.003$, in $2D$, and $d_{BB}=2.50\pm0.02$, in $3D$.
        \label{fig::snap}}
\end{figure*}

In Fig.~\ref{fig::snap} we show the evolution of the fracture line on a
large ranked surface with the fraction of occupied bonds $p$. As
displayed, the lattice is initially seeded by a set of disconnected
bridge elements at low values of $p$, while for $p \rightarrow 1$ the
fracture finally emerges towards a singly connected line that divides
the system in two. As we show later in this article, our results reveal
that this line is fractal with dimension $d_{BB} \approx 1.22$.
Interestingly, this value is statistically identical to the dimensions
of fractures generated from different models previously investigated
\cite{Cieplak94,Cieplak96,Fehr09,Fehr11}. However, at the percolation
threshold value of the classical random percolation model
\cite{Broadbent57,Stauffer94}, $p=p_c$, the set of bridge bonds
appearing in any configuration of our model should be identical to the
set of the so-called {\it anti-red bonds} in random percolation
\cite{Coniglio89}. As first proposed by Coniglio \cite{Coniglio89} and
numerically verified by Scholder \cite{Scholder09}, at $p=p_c$, this set
is also fractal, but with dimension $1/\nu$ ($=3/4$ in 2D), where $\nu$
is the correlation length exponent.

Motivated by this substantial change in the fractal behavior at
distinct stages of our fracturing process, in what follows we address
how the set of bridges scales with the fraction of occupied bonds and
the system size at $p=p_c$ analogously to a theta point
\cite{deGennes79,Chang91,Poole89}, while for all values of $p$ above
$p_c$, it has a fractal dimension $d_{BB}$. Moreover, we introduce a new
tricritical crossover exponent, which we study up to dimension six, the
upper-critical dimension of percolation.
\section{Results}
\begin{figure}
      \begin{center}
        \includegraphics[width=\columnwidth]{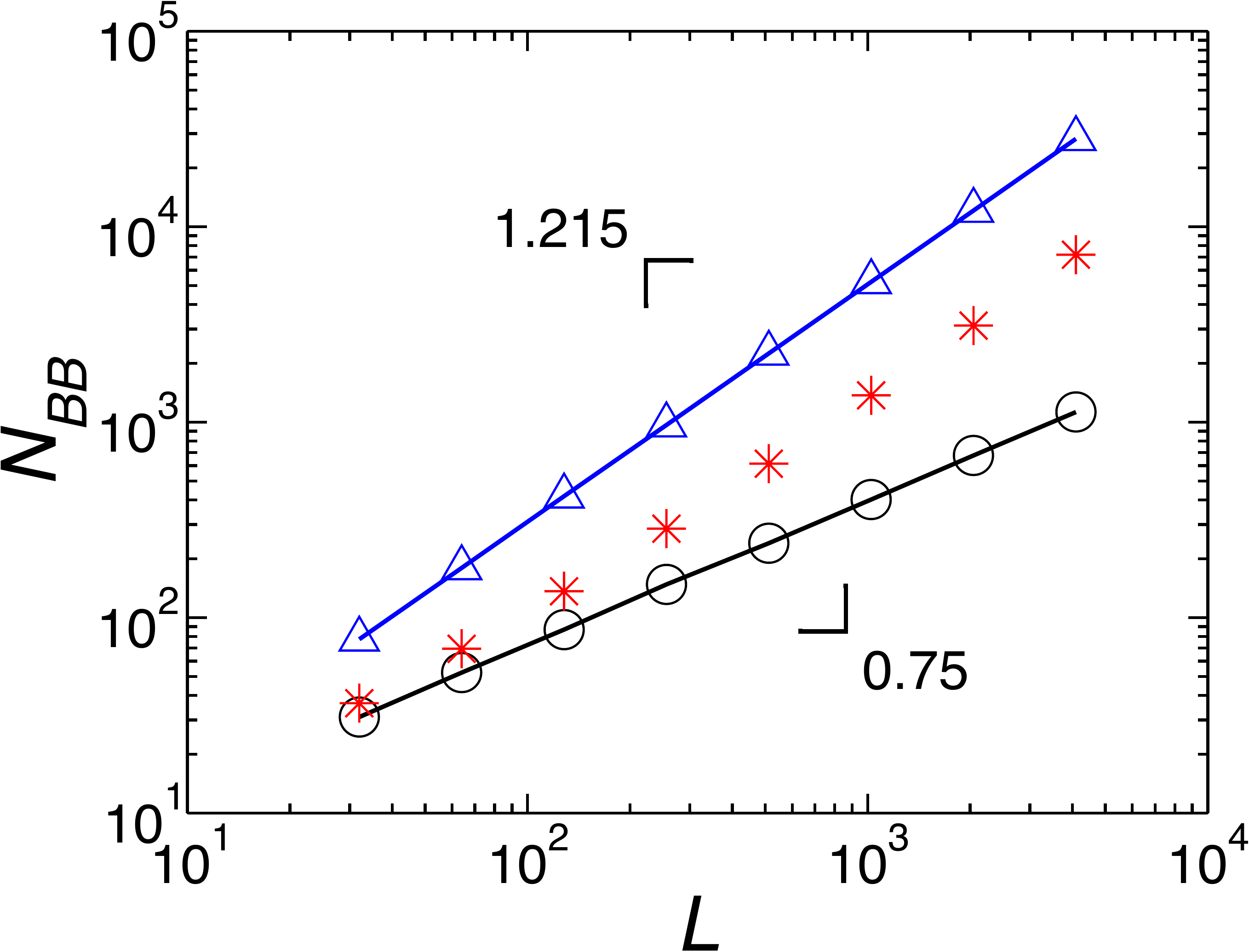}
      \end{center}
      \caption{
        Crossover in the size dependence.
        Number of bridge bonds, $N_{BB}$, for different $p$, namely, $p=p_c=0.5$ (circles), $p=0.51$ (stars), and $p=0.8$ (triangles).
        The solid lines stand for the best fit.
        At $p=p_c$, as conjectured by Coniglio \cite{Coniglio89}, $N_{BB}\sim L^{1/\nu}$, where $\nu=4/3$ is the critical exponent of the correlation length in $2D$.
        For $p>p_c$, the number of bridge bonds scales with $L^{d_{BB}}$.
        A crossover between the two regimes in system size is observed (stars) for $p$ in the neighborhood of $p_c$.
        Systems of size $L^2$ have been considered, with $L$ ranging from $32$ to $4096$.
        All results have been averaged over $10^4$ samples.
        Error bars are smaller than the symbol size.
        \label{fig::mbbsize}}
\end{figure}

\begin{figure}
      \begin{center}
        \includegraphics[width=\columnwidth]{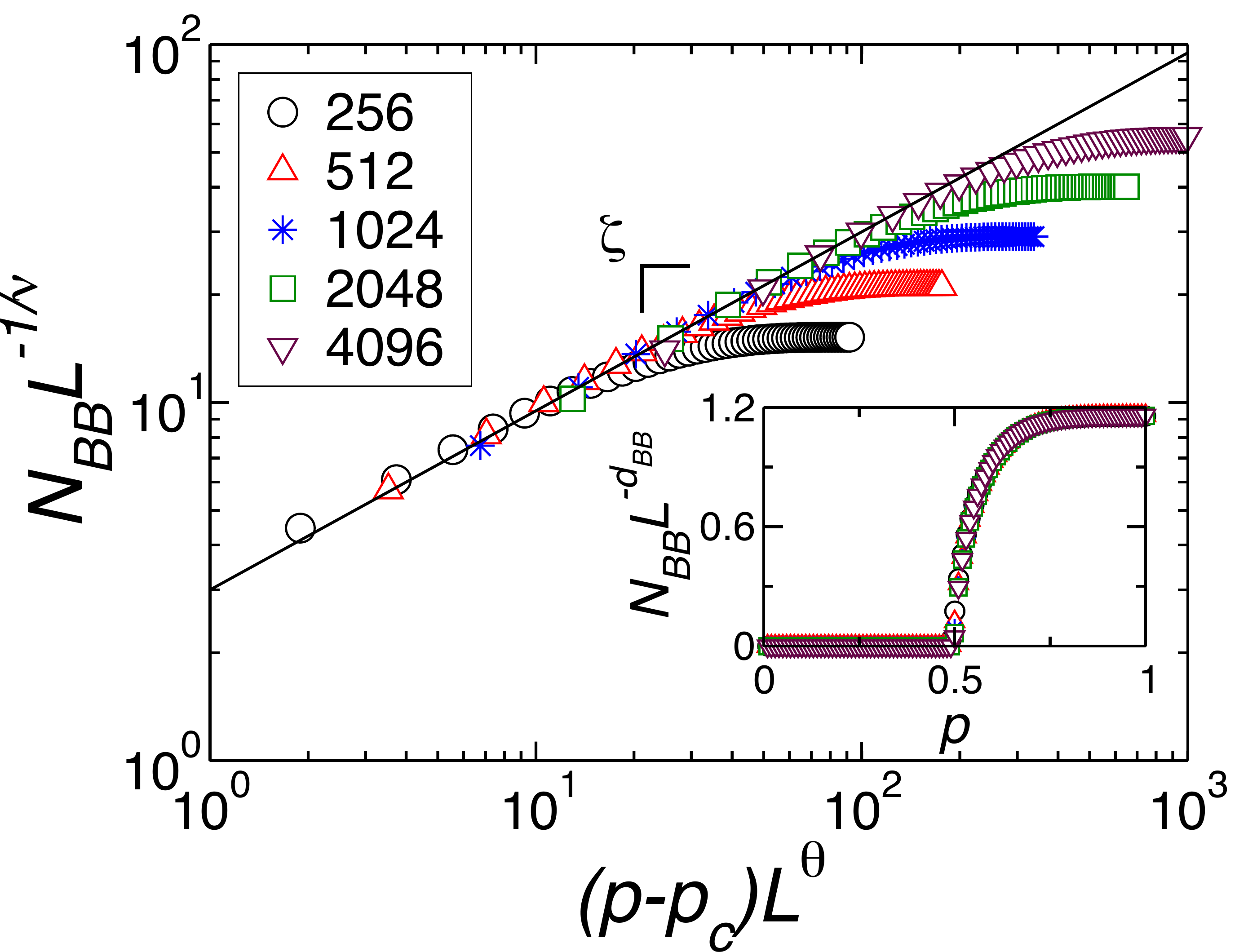}
      \end{center}
      \caption{
        Tricritical scaling, crossover, and data collapse.
        Number of bridge bonds, $N_{BB}$, as a function of the fraction
of occupied bonds, $p$, for $2D$ with different system sizes $L=\{256,512,1024,2048,4096\}$.
        The scaling function given by equation~(\ref{eq::scaling}) is applied, with $\theta=0.93$, obtaining $\zeta=0.50\pm0.03$.
        In the inset, $N_{BB}$ has been rescaled by $L^{d_{BB}}$, where $d_{BB}=1.215$.
        All results have been averaged over $10^4$ samples.
        \label{fig::bbscale}}
\end{figure}

We performed simulations of fracturing ranked surfaces on square
lattices. It is worth noting that, despite the similarities with random
percolation, the suppressing of connectivity poses a statistically
different problem.  For example, while for $p\rightarrow1$ there is only
a single configuration in random percolation (all bonds occupied), in
ranked percolation there are $N!$, evenly weighted, possible
configurations, where $N$ is the total number of bonds.  In classical
percolation the total number of configurations is $2^N$.
Figure~\ref{fig::mbbsize} shows the dependence of the number of bridge
bonds $N_{BB}$ on system size, for different fractions of occupied
bonds, namely, $p=p_c=0.5$, $p=0.51$, and $p=0.8$.  As expected, at the
percolation threshold of classical percolation ($p=p_c$), the number of
bridge bonds diverges with system size as $N_{BB}\sim L^{1/\nu}$, where
$\nu$ is the correlation length exponent, with $\nu=4/3$ in $2D$; while
for $p=0.8$, $N_{BB}\sim L^{d_{BB}}$, with $d_{BB}=1.215\pm0.003$.  The
latter is in fact observed at any $p>p_c$.  We found the same result
(see {\it Appendix}) for site percolation and on other
lattices (star, triangular, and honeycomb), which provides strong
evidence for the universality of this exponent.  For $p\gtrsim p_c$,
like $p=0.51$ we can observe, as depicted in Fig.~\ref{fig::mbbsize}, a
crossover between the two different regimes.  The inset of
Fig.~\ref{fig::bbscale} shows $N_{BB}$, rescaled by $L^{d_{BB}}$, as a
function of $p$, for different system sizes.  The number of bridge bonds
grows with $p$, such that, $N_{BB}\sim(p-p_c)^\zeta$, where
$\zeta=0.50\pm0.03$ is a novel exponent, which we call bridge-growth
exponent.  The overlap of the different curves confirms that the fractal
dimension of the bridge bonds above $p_c$ is $d_{BB}$, for all $p>p_c$.
This result differs from classical percolation where fractality is
solely observed at criticality \cite{Broadbent57,Stauffer94} while,
above $p_c$, bridge bonds are only observed for finite systems
\cite{Coniglio89}.

For polymer chains, at high temperatures, the excluded volume prevails
over attractive forces and the chain can be described as a self-avoiding
walk. When the temperature is reduced, the attractive forces become
relevant leading, at a theta-temperature, to a new exponent at
the crossover between two dimensions \cite{deGennes79,Chang91,Poole89}.
Analogously, in ranked percolation the fractal dimension of the bridge
bonds is $1/\nu$, at $p=p_c$, between $d_{BB}$ above $p_c$ and zero
below $p_c$.  For the tricritical scaling we verify the following
\textit{ansatz},
  \begin{equation}\label{eq::scaling}
    N_{BB}=L^{1/\nu}\mathcal{F}\left[\left(p-p_c\right)L^\theta\right],
  \end{equation}
where $\mathcal{F}[x]\sim x^{\zeta}$ for $x\neq0$, and is nonzero at
$x=0$; and the exponent $\theta$ is the crossover exponent.  Therefore,
the following relation is obtained,
  \begin{equation}\label{eq::scaling.relation}
    \theta=\zeta^{-1}\left(d-\varphi-\frac{1}{\nu}\right),
  \end{equation}
where $\varphi=d-d_{BB}$.  In the main plot of Fig.~\ref{fig::bbscale}
we see, for $2D$, the scaling given by equation~(\ref{eq::scaling}),
with $\theta=0.93$.

The results above disclose a tricritical $p_c$ below which the fraction
of bridges in the bridge line vanishes in the thermodynamic limit. This
is identical to a minimum height in the bridge line, $H=h_\mathrm{min}$,
in the context of landscapes (Figs.~\ref{fig::rank}(a)-(b)). For a
cumulative distribution function of heights $H$, $P(H\leq h)$, the
minimum is given by $P(H\leq h_\mathrm{min})=p_c$.  Note that, for
uncorrelated landscapes, regardless the distribution of heights, the set
of bridges only depends on their position in the rank.  To observe the
new set of exponents on discretized landscapes
(Fig.~\ref{fig::rank}(c)), $N_{BB}$ is the number of sites in the bridge
line with both neighbors (one at each side) having height lower than
$h$, where $P(H\leq h)=p$.

\begin{figure}
      \begin{center}
        \includegraphics[width=\columnwidth]{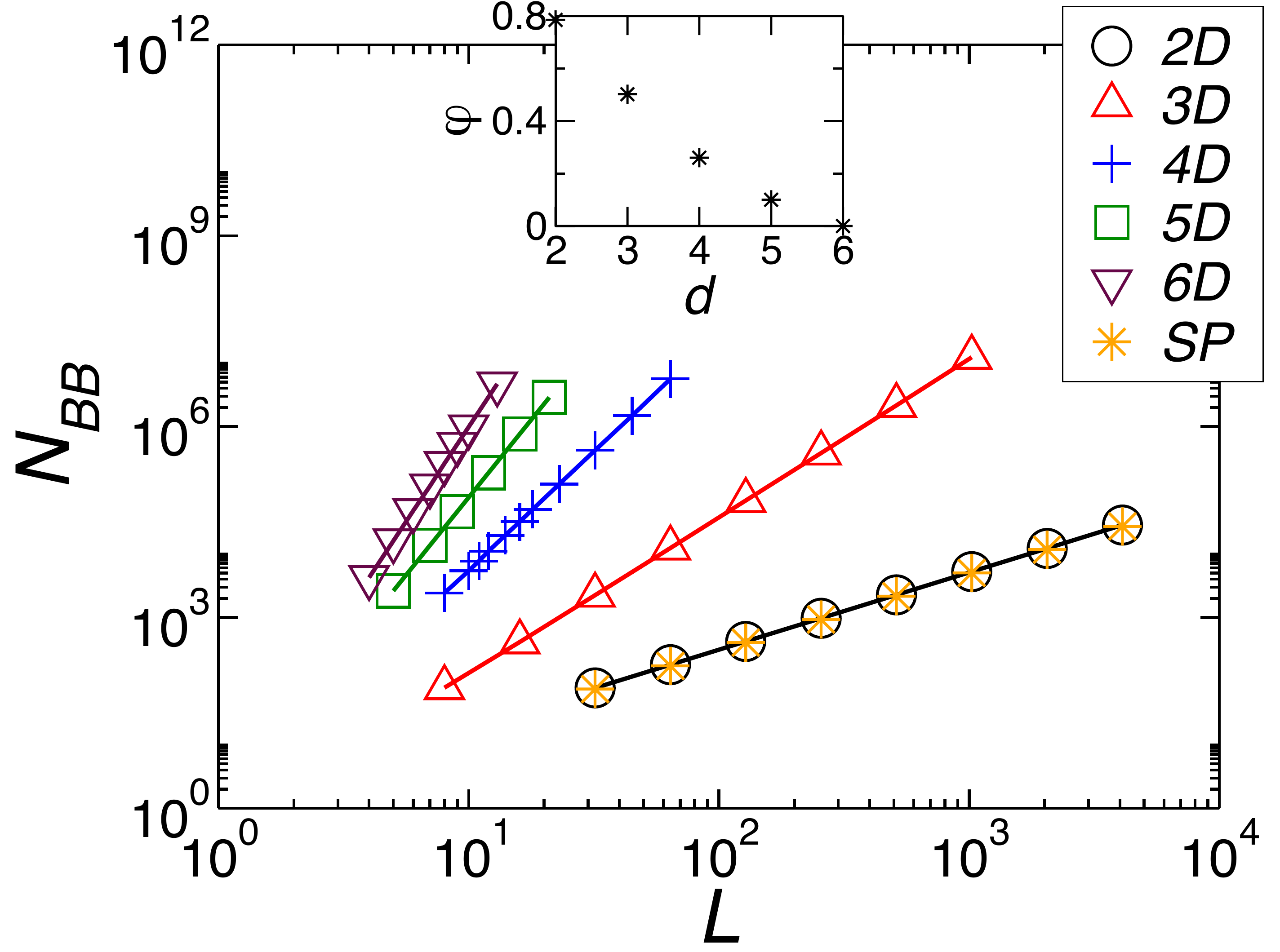}
      \end{center}
    \caption{
      Dependence on the spatial dimension and approach to the
mean-field limit. 
      Size dependence of the number of bridge bonds, $N_{BB}$, in
      the limit $p=1$, up to dimension six.  The solid lines stand for the
      best fit.  The mass of the shortest path in loopless percolation (SP) in
      $2D$ is also included for comparison.  Results have been averaged over
      $10^4$ samples for $2D$ and $3D$, and $10^2$ samples for higher
      dimensions.  Error bars are smaller than the symbol size.  The inset
      shows the dependence on the spatial dimension of the exponent $\varphi$.
      At the upper-critical dimension of percolation, $d=6$, the set of bridge
      bonds becomes dense having the spatial dimension.
      \label{fig::nbbdim}} 
\end{figure}

To study the dependence of exponents $\zeta$ and $\varphi$ on the
spatial dimension, we analyze the same problem up to dimension six.  On
a simple-cubic lattice ($3D$), above $p_c$, the set of all bridge bonds
has a fractal dimension $d_{BB}=2.50\pm0.05$ and grows with
$\zeta=1.0\pm0.1$ (see {\it Appendix}).
Figure~\ref{fig::nbbdim} shows the size dependence of $N_{BB}$, in the
limit $p=1$, for lattices with size $L^d$, where $2\leq d\leq 6$ is the
spatial dimension.  In the inset, we plot the exponent $\varphi$
as a function of $d$.  Since the set of bridge bonds blocks connectivity
from one side to the other, its fractal dimension must
follow $d-1\leq d_{BB} \leq d$, i.e., $0\leq\varphi\leq 1$.  With
increasing dimension $\varphi$ decreases.  At the upper-critical
dimension of percolation, $d_c=6$, $\varphi=0.0\pm0.1$ and the set of
bridge bonds becomes dense having the spatial dimension $d$.
Table~\ref{tab::exponents} summarizes the exponents for dimensions $2$
to $6$.  The bridge-growth exponent grows with $d$ and converges to
$\zeta=1.5\pm0.7$ at the upper-critical dimension.  For $d>6$, above the
critical dimension of percolation, the exponent $\varphi$ remains zero
and the dimension of the set of bridges is equal to $d$.  This can be
understood from the fact that above $d_c$ one has an infinity of
spanning clusters \cite{Newman81} and thus many more possible bridges.  Since the
dimension of the set of bridges at $p=p_c$ is $1/\nu$ and above $p_c$ is
$d$, the crossover exponent increases with $d$, with the relation given
by equation~(\ref{eq::scaling.relation}), where $\varphi=0$.
\begin{table}
  \caption{ 
    Table of exponents.
    Values of the exponents $\zeta$ and
    $\varphi$ up to dimension six.  With increasing dimension, the $\zeta$
    exponent converges to $\zeta=1.5\pm0.7$ and $\varphi$ goes to zero,
    revealing that the set of bridge bonds is dense.
    \label{tab::exponents} } 
  \begin{tabular}{ccc}
    \hline\hline $d$ & $\zeta$ & $\varphi$ \\
    \hline $2$ & $0.50\pm0.03$ & $0.785\pm0.003$ \\
    $3$ & $1.0\pm0.1$ & $0.50\pm0.02$ \\
    $4$ & $1.3\pm0.5$ & $0.26\pm0.08$ \\
    $5$ & $1.4\pm0.6$ & $0.1\pm0.2$ \\
    $6$ & $1.5\pm0.7$ & $0.0\pm0.1$ \\
    \hline 
  \end{tabular} 
\end{table}

Since one can interchange occupied and empty bonds, there exists a
symmetry between bridges and cutting bonds (red bonds), so one can raise
the question of what happens when bonds are removed from a percolating
system with the constraint that connectivity cannot be broken.
Initially all bonds are occupied and at the end, since cutting bonds are
never removed, a line of cutting bonds is obtained, which we denote here as
the {\it cutting line}.  Above $p_c$, as in the classical case, the percolation
cluster is compact and there are no cutting bonds.  For $p<p_c$, the set
of cutting bonds is fractal with the same fractal dimension as the
bridge-bond set, $d_{CB}=d_{BB}$, whereas at $p_c$ it is $1/\nu$.  For
the crossover scaling a similar \textit{ansatz} to the one given by
equation~(\ref{eq::scaling}) is verified, where the argument of the
scaling function is then $(p_c-p)L^\theta$.  The same hyperscaling of
equation~(\ref{eq::scaling.relation}) is obtained with
$\varphi=d-d_{CB}$.  At $d=2$, our numerical results corroborate the
hypothesis of the same $\zeta$ and $\varphi$, for cutting and bridge
bonds (see {\it Appendix}).  For $d>2$, the set of
cutting bonds is a line with $d_{CB}\leq2$ and the one of bridge bonds
has a dimension above $d-1$, so that the fractal dimensions differ.
Above the critical dimension, the set of cutting bonds has dimension
two, like the shortest path at $p_c$ \cite{Havlin84}, for any $p\leq
p_c$.

Fisher \cite{Fisher61} proposed a bond-site transformation to map bond
percolation on a lattice onto site percolation on a covering graph. For
example, the covering graph of the square lattice is obtained by adding
all diagonal edges to every other face \cite{Wierman95}. Considering
such a mapping for ranked percolation and, following the analogy with a
random landscape, where sites are sequentially removed from the lowest
number (height) and suppressing global disconnection, the obtained
cutting line is identical to the bridge one with the constraint applied
in the perpendicular direction. The same is also observed for
site-ranked percolation on a square lattice. Given the relation between
connectivity and disconnection, in both cases, the cutting version needs
to be defined on different topologies, namely, the star lattice for the
square (on sites) and the same lattice for its covering graph, but with
swapped cells of diagonal bonds. For the triangular lattice, the
relation between cutting and bridges sites is straightforward without
requiring additional connections.
\begin{figure}
      \includegraphics[width=\columnwidth]{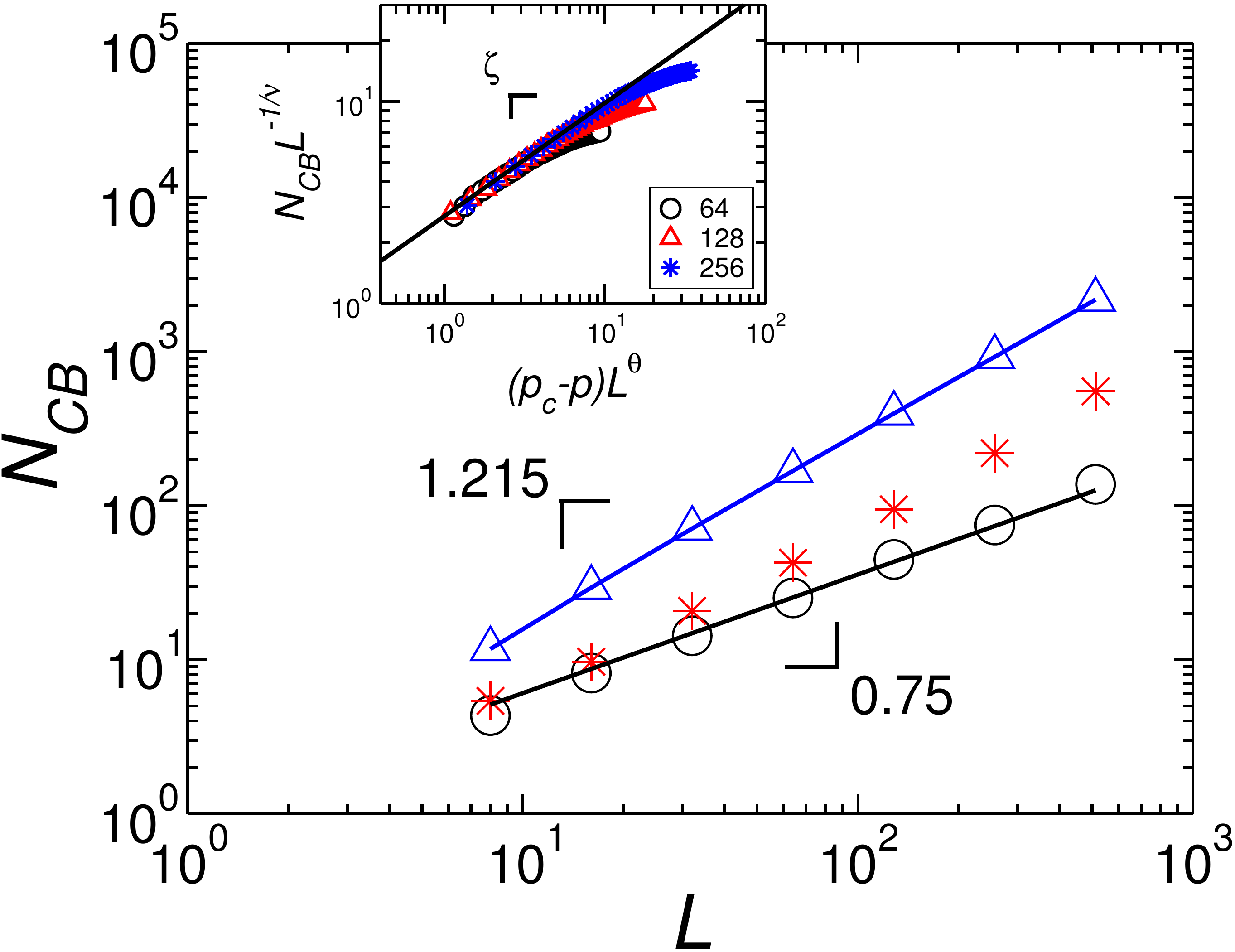}
      \caption{
        Following the rank, bonds are sequentially removed except
        the cutting bonds.
        Size dependence of the number of cutting bonds, $N_{CB}$, for
        different fractions of occupied bonds $p$, namely
        $p=p_c=0.5$ (circles), $p=p=0.49$
        (stars), and $p=0.2$ (triangles). The solid lines are guides to the
        eye. At $p=p_c$, $N_{CB} \sim
        L^{1/\nu}$, where $\nu = 4/3$ is the critical exponent related with
        the correlation length. For $p<p_c$, the number of cutting bonds
        scales with $L^{d_{CB}}$. A crossover between the two regimes with the
        system size is observed (stars) for $p$ in the neighborhood
        of $p_c$. Systems of size $L^2$ have been considered, with $L$ ranging
        from $8$ to $512$. Results have been averaged over $10^3$ samples for
        most system sizes and over $50$ samples for the largest one. Error
        bars are smaller than the symbol size.
        The inset shows the number of cutting bonds, $N_{CB}$, as a function
        of occupied bonds, $p$, for two-dimensional lattices with $L$ ranging
        from $64$ to $256$ (averaged over $10^2$ samples).
        The scaling is given by Eq.~($1$) in the article, where the argument is
        then $(p_c-p)L^\theta$, with $\theta=0.93$.
        We obtain $\zeta=0.56\pm0.08$, which is compatible with the one for bridge bonds.
        \label{fig::cb}}
\end{figure}
\section{Discussion}
For several different models in 2D the same fractal dimension as the one
of bridge bonds has been reported. Here we discuss how some of them can
be related with the process of fracturing ranked surfaces. Let us
construct ranked percolation configurations on a random landscape
starting with the bond with the largest number (height) and then
occupying bonds sequentially in order of decreasing number. Each time a
chosen bond closes a loop, it is not occupied (loopless). One stops the
procedure when, for the first time, a path of connected sites spans from
one side to the other (percolation threshold). Since all clusters are
trees, the backbone and the shortest path are the same and their fractal
dimension is identical to the bridge-bond line, as verified in
Fig.~\ref{fig::nbbdim}. In fact, given the connectivity/disconnection
relation discussed in the previous section, this line corresponds to the
bridge line when bonds are occupied from the lowest to the largest
number in the covering graph.  This line also corresponds to the line of
cutting bonds when bonds are removed from the smallest to the largest
number. A similar procedure was also proposed to obtain the minimum
spanning tree (MST), for which the same fractal dimension is found for
the paths between all pairs of sites \cite{Dobrin01}. There bonds are
sequentially removed (from the highest to the lowest value) under the
constraint that all sites remain connected. If the constraint is relaxed
such that only connectivity between two opposite borders is imposed, the
cutting-bond line is obtained. Therefore, the cutting-bond line is the
path between borders on the MST for which the largest value (height) is
minimum, a valid relation in any spatial dimension $d$.
\begin{figure}
      \includegraphics[width=\columnwidth]{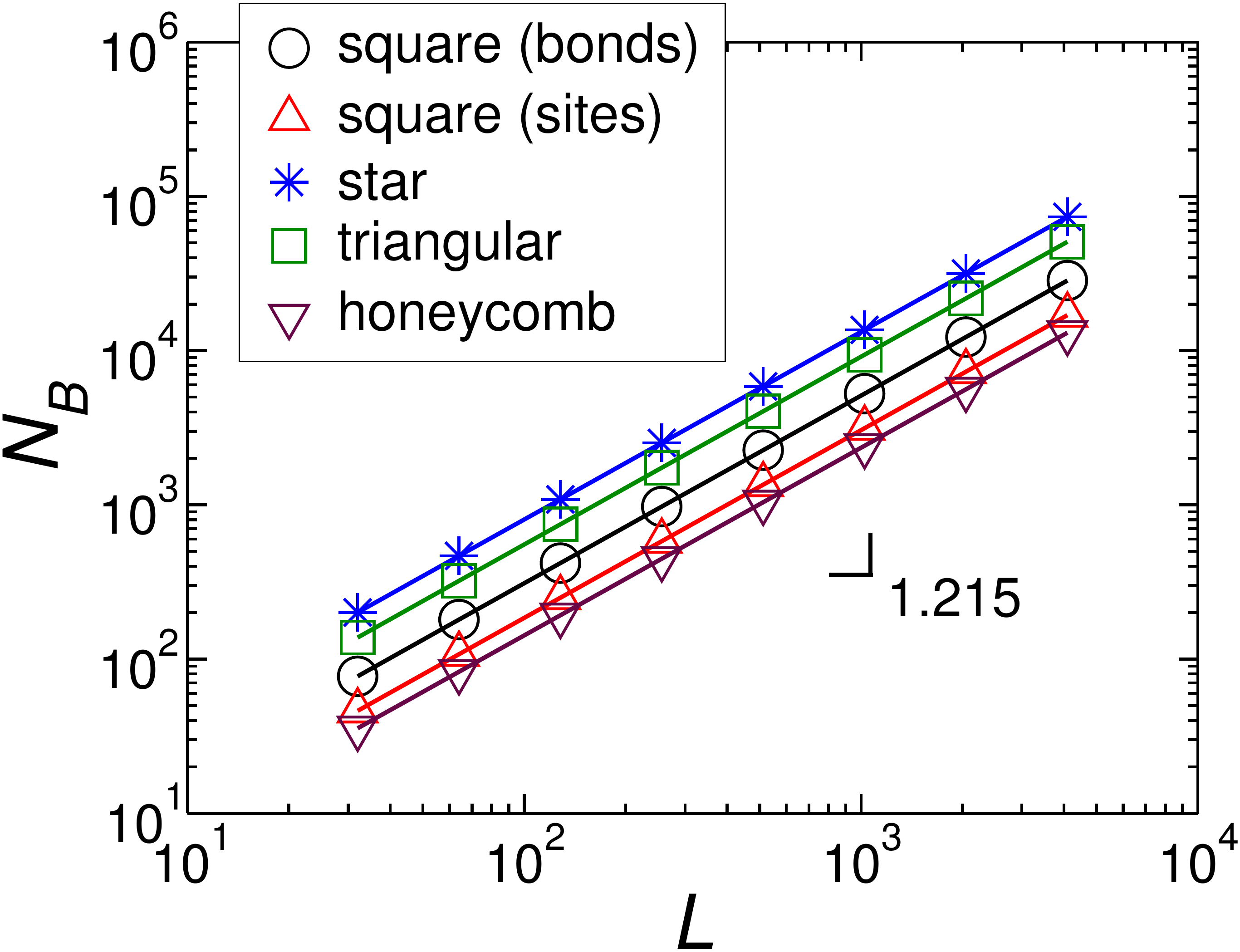}
      \caption{
      Size dependence of the number of bridges, $N_B$, on several different lattices.
      The solid lines are guides to the eye, all with slope $1.215$.
      The best fit to each data set gives the same fractal dimension of $1.215\pm0.005$.
      Systems of linear size $L$ have been considered, with $L$ ranging from $32$ to $4096$.
      Results have been averaged over $10^5$ samples.
      \label{fig::difnet}}
\end{figure}
\begin{figure}
      \includegraphics[width=\columnwidth]{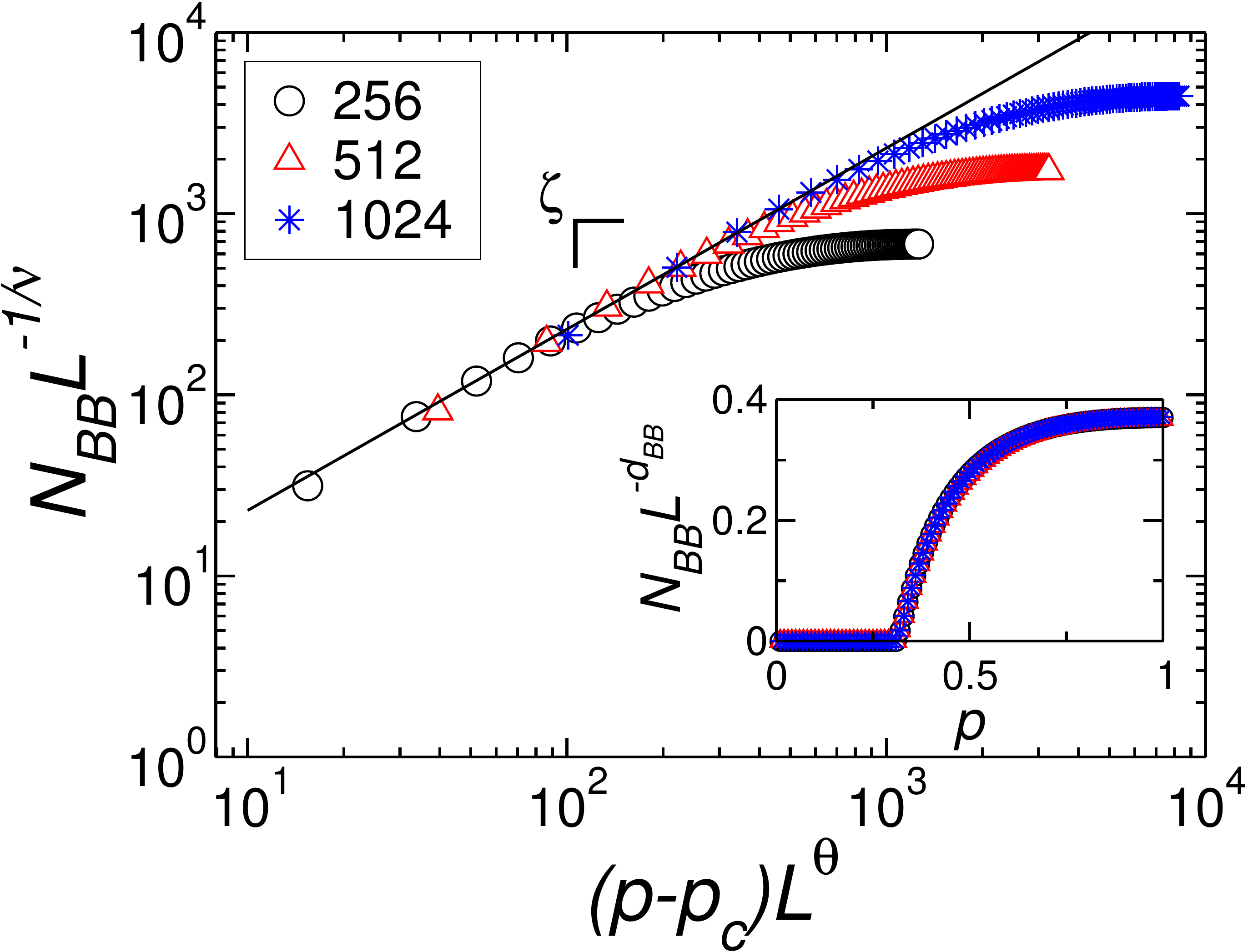}
      \caption{
        Number of bridge bonds, $N_{BB}$, as a function of the fraction
of occupied bonds, $p$, for $3D$ (simple-cubic lattices) with different sizes $L=\{256,512,1024\}$.
        The scaling function proposed in this article is applied, with $\theta=1.36$, obtaining $\zeta=1.0\pm0.1$.
        In the inset, $N_{BB}$ has been rescaled by $L^{d_{BB}}$, where $d_{BB}=2.50$.
        All results have been averaged over $10^4$ samples.
        \label{fig::bbscale3d}}
\end{figure}

The fractal dimension $d_{BB}$ was also observed for the backbone of the
optimal path crack (OPC) \cite{Andrade09,Oliveira11}. For a random
landscape, the sequence of optimal paths between opposite borders is
obtained and their highest site removed. Every such path crosses the
bridge line. The highest site on successive paths is either on the
bridge line itself or is higher than the lowest bridge-line site.
Therefore, as in the loopless percolation described before, the removed
sites percolate when all sites on the bridge line are removed, which is
the backbone of the crack, giving the same fractal dimension $d_{BB}$.
This is in fact the case for the crack of any sequence of self-avoiding
paths between opposite borders. Since in this case the duality between
cutting and bridges is not used (only valid in 2D), the relation between
OPC and bridge bonds is still true in higher dimensions.

Let us now define the optimal minimax path (OMP) in the following way.
One starts selecting the set of paths on the landscape for which the
highest-ranked site is minimum (minimax paths). Such set is then reduced
to include only the paths for which the second highest site is also
minimum and one proceeds iteratively to the following sites, until a
unique path is obtained. This path is the optimal path in strong
disorder \cite{Porto97,Porto99}, since under such disorder strength each
site is higher than the sum over the height of all sites with lower
rank. This path is also identical to the backbone of the
loopless-ranked-percolation cluster when occupied from the lowest to the
largest number and, therefore, the cutting line in any dimension.

In summary, suppressing connectivity (disconnection) between opposite
borders on ranked surfaces leads to a fractal set of bridge (cutting)
bonds, with a universal fractal dimension, even far away from the critical
point of classical percolation. In $2D$, there is an
equivalence between bridges and cutting bonds and $d_{BB}=d_{CB}$,
whereas for $d>2$ cutting bonds are still a fractal line but bridge
bonds form a surface. The discussed models are then split into two groups. The
ones suppressing connectivity (e.g., watershed \cite{Cieplak94,Fehr09}
and optimal path crack \cite{Andrade09,Oliveira11}) are in the bridges
universality class, while the ones keeping connectivity (e.g., optimal
path on the MST or in strong disorder media \cite{Porto97}) are in the
universality class of cutting bonds. For $d>6$, $d_{BB}=d$ and
$d_{CB}=2$, so we conjecture that the upper-critical dimension of
bridges and cutting bonds is also $d_c=6$. Finally, we show that, at the
percolation threshold of classical percolation, ranked percolation
displays a theta-point-like crossover.
\begin{figure}
      \includegraphics[width=\columnwidth]{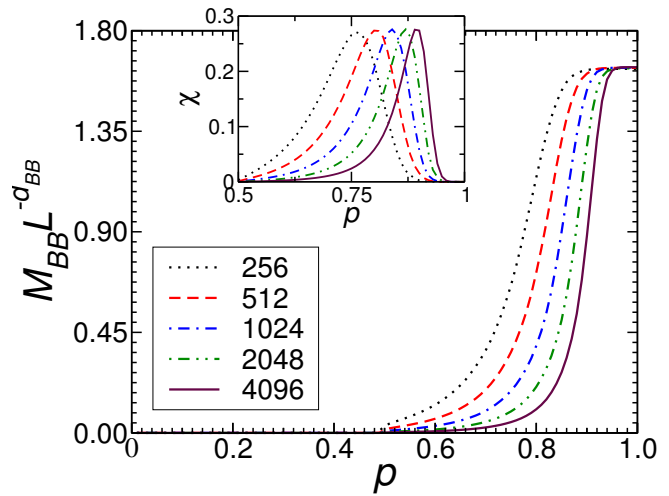}
      \caption{
        Size of the largest cluster of sites on the edge of bridge bonds, $M_{BB}$, rescaled by $L^{d_{BB}}$, as a function of the fraction of occupied bonds $p$.
        The inset shows $\chi$ defined by Eq.~(\ref{eq::def.chi}).
        Results have been obtained on a square lattice with $L$ ranging from $256$ to $4096$.
        All results have been averaged over $10^4$ samples.
        \label{fig::bbp}}
\end{figure}

This work opens up several challenges. Besides the need for a more
precise numerical estimation of the bridge-growth exponent, it would be
interesting to formulate a renormalization group scheme and to obtain
the new set of exponents from analytic treatments such as, e.g., exact
results in the mean-field limit and a Schramm-Loewner evolution in two
dimensions. Another interesting possibility is to find the corresponding
exponents in other related problems with different universality classes
like, e.g., the Kasteleyn-Fortuin clusters in the $q$-state Potts model
\cite{Coniglio89,Fortuin72,Wu78,Coniglio80,Deng04} with or without
magnetic field.  Regarding the fractal dimension of the surface of
discontinuous percolation clusters \cite{Araujo10,Schrenk11}, it would
also be interesting to understand how it relates with the herein
introduced bridge-bonds universality class. For the cutting bonds, the
study of the crossover for higher dimensions represents another computational
challenge.  Finally, it would also be interesting to try to identify the
third scaling field of our theta-like point.
\section{Methods}
All numerical results have been obtained with Monte Carlo simulations.
Results have been averaged over $10^4$ samples for $2D$ and $3D$, and
$10^2$ samples for higher dimensions.
\section{Author contributions}
K.J.S. and N.A.M.A. carried out the numerical experiments. All authors
conceived and designed the research, analyzed the data, worked out the
theory and wrote the manuscript.
\begin{acknowledgments}
We acknowledge financial support from the ETH
Risk Center.  We also acknowledge the Brazilian agencies CNPq, CAPES and
FUNCAP, and the Pronex grant CNPq/FUNCAP, for financial support.
\end{acknowledgments}
\appendix
\section{Cutting bonds}
In Fig.~\ref{fig::cb} we see the size dependence of the number of cutting bonds in $2D$ when bonds are removed with the constraint that cutting bonds are never removed.
As discussed in the article for bridge bonds, a crossover is observed at $p=p_c$, between two fractal dimensions.
In $2D$, for $p<p_c$ the fractal dimension of the cutting bonds $d_{CB}$ is the same as observed for the bridge bonds for $p>p_c$, whereas at $p_c$ it is $1/\nu$.
\section{Universality of the fractal dimension of bridge bonds}
In Fig.~\ref{fig::difnet} one sees results for the bridge-bond line in several different lattices, namely, square, star, triangular, and the honeycomb lattices.
The data points for bridge sites on the square lattice are also included.
For all considered lattices the obtained fractal dimension of the bridge line is $1.215\pm0.005$, a strong evidence for the universality of this fractal dimension.
\section{Crossover scaling in $3D$}
Figure~\ref{fig::bbscale3d} shows the crossover scaling for the number of bridge bonds $N_{BB}$ on a simple-cubic lattice with different sizes.
Bonds are occupied except for the bridge bonds.
We applied the scaling function proposed in Eq.~(1) of the article with $\theta=1.36$, giving $\zeta=1.0\pm0.1$.
\section{Percolation of bridges}
As discussed in the article, the set of bridge bonds merges towards a single connected line.
In Fig.~\ref{fig::bbp} we see the size of the largest cluster of sites on the edge of bridge bonds $M_{BB}$, rescaled by $L^{d_{BB}}$, as a function of $p$, for different system sizes.
The inset shows $\chi$, defined as,
\begin{equation}\label{eq::def.chi}
\chi=\frac{1}{L^{2d_{BB}}}\left[\sum_i s^2_i - M_{BB}^2\right],
\end{equation}
where the sum runs over all clusters of sites on the edge of bridge bonds and $s_i$ is their size.
Alike percolation on a line, in the thermodynamic limit, the bridge-bond line only percolates at $p\rightarrow1$.
\end{document}